\documentclass{aa}
\usepackage{psfig}
\usepackage{amssymb}
\psfigurepath{fig}
\usepackage{graphicx,natbib}
\usepackage{amsmath,epsfig,rotating,amssymb}
\usepackage{ulem}
\usepackage{amsmath}
\begin{document}
\title{Mid infrared properties of distant infrared luminous galaxies
\thanks{Based on observations collected with the Spitzer Space Telescope 
(which is operated by the Jet Propulsion Laboratory, California Institute of 
Technology under NASA contract 1407) and on observations with ISO 
(an ESA project with instruments funded by ESA Member States 
(especially the PI countries: France, Germany, the Netherlands and 
the United Kingdom) with the participation of ISAS and NASA)} }

\author{\bf D.~Marcillac\inst{1}, D.~Elbaz\inst{1}, R.R.~Chary\inst{2}, M.~Dickinson\inst{3}, F. Galliano\inst{4} and G.~Morrison\inst{3,5}}
\offprints{D. Marcillac, email: marcilla@cea.fr, dmarci@as.arizona.edu}
\institute{DSM/DAPNIA/Service d'Astrophysique, CEA/SACLAY, 91191 Gif-sur-Yvette Cedex, France
\and Spitzer Science Center, California Institute of Technology, MC 220-06, Pasadena, CA 91125, USA
\and National Optical Astronomy Observatory, 950 North Cherry Street, Tucson, AZ 85719, USA
\and Observational Cosmology Laboratory, NASA Goddard Space Flight Center, Greenbelt MD 20771, USA
\and Institute for Astronomy, University of Hawaii, and Canada-France-Hawaii Telescope Kamuela, Hawai 96743}
%

\titlerunning{Mid infrared properties of distant infrared luminous  galaxies}
\authorrunning{Marcillac et al.}

\abstract{We present evidence that the mid infrared ( MIR , rest frame 5-30 $\mu$m ) is a good tracer of the total infrared luminosity, L(IR)($=~L[8-1000\,\mu{\rm m}]$), and star formation rate (SFR), of galaxies up to $z\sim$ 1.3. We use deep MIR images from the Infrared Space Observatory (ISO) and the Spitzer Space Telescope in the Northern field of the Great Observatories Origins Deep Survey (GOODS-N) together with VLA radio data to compute three independant estimates of L(IR). 
The L(IR,MIR) derived from the observed 15 and/or 24\,$\mu$m flux densities using a library of template SEDs, and L(IR,radio), derived from the radio (1.4 and/or 8.5 GHz) using the radio-far infrared correlation, agree with a 1-$\sigma$ dispersion of 40\,\%.
We use the k-correction as a tool to probe different parts of the MIR spectral energy distribution (SED) of galaxies as a function of their redshift and find that on average distant galaxies present MIR SEDs very similar to local ones. However, in the redshift range $z=$ 0.4-1.2, L(IR,24\,$\mu$m) is in better agreement with L(IR,radio) than L(IR,15\,$\mu$m) by 20\,\%, suggesting that the warm dust continuum is a better tracer of the SFR than the broad emission features due to polycyclic aromatic hydrocarbons (PAHs). We find marginal evidence for an evolution with redshift of the MIR SEDs: two thirds of the distant galaxies exhibit rest-frame MIR colors (L(12\,$\mu$m)/L(7\,$\mu$m) and L(10\,$\mu$m)/L(15\,$\mu$m) luminosity ratios) below the median value measured for local galaxies. Possible explanations are examined but these results are not sufficient to constrain the physics of the emitting regions. If confirmed through direct spectroscopy and if it gets amplified at higher redshifts, such an effect should be considered when deriving cosmic star formation histories of dust-obscured galaxies. We compare three commonly used SED libraries which reproduce the color-luminosity correlations of local galaxies with our data and discuss possible refinements to the relative intensities of PAHs, warm dust continuum and silicate absorption.
}
\maketitle
\section{Introduction}
%
Luminous infrared phases, during which galaxies radiate the bulk of their light in the infrared (IR) and are producing stars at rates larger than 20 M$_{\odot}$ yr$^{-1}$, play a dominant role in the cosmic star formation history (CSFH) and in the production of the cosmic infrared background (CIRB) 
 \citep{CE,E02,Elbaz99,xu03,fran01,fran03,L04,chary04,papovich04}. Those studies suggest that at least half of present-day stars were formed in ``infrared luminous galaxies'', which we define here as the combination of luminous (LIRGs, $10^{11}$L$_{\sun}\leq L(IR)<10^{12}$L$_{\sun}$) and ultra-luminous (ULIRGs, $L(IR)\geq 10^{12}~$L$_{\sun}$) infrared galaxies. Indeed, the comoving density of mid (15 \& 24\,$\mu$m) and total IR luminosities ($L(IR)=~L[8-1000\,\mu{\rm m}]$) due to infrared luminous galaxies was about 70 times larger at $z\sim$ 1 than it is today \citep{Lefloch,CE}. However, existing instruments do not allow direct measurement of $L(IR)$ and one must rely on correlations between other wavelength ranges and $L(IR)$ to derive its value as well as the obscured star formation rate (SFR) that it quantifies \citep{Kennicutt98}. 

$L(IR)$ can be derived from a single measurement in the mid IR (MIR, rest frame 5-30 $\mu$m) for local galaxies with a 68\,\% uncertainty of 30-40\,\% depending on the MIR wavelength (see Figs.1\&2 from \cite{CE}). Similarly, the radio continuum exhibits an impressive correlation with L(IR) over four orders of magnitude with an uncertainty of only 5\,\% over nearly 2000 galaxies \citep{Yun01}. However, it must be noted that when limited to the 162 infrared luminous galaxies of the sample of \cite{Yun01}, this dispersion rises to 46\,\% which is comparable to the one observed between the mid-infrared and total IR luminosities. The origin of both correlations, linking the radio and MIR to L(IR), is not yet completely understood since several physical processes are generally advocated to explain the emission in both domains. 

It is generally acknowledged that the radio emission is produced by synchrotron radiation from relativistic electrons and free-free emission from HII regions. Most of the energy required for the acceleration of electrons is produced by the supernova remnants of stars more massive than 8 M$_{\odot}$, which also dominate the ionization in H II regions. However, whether this is enough to explain the extent of the correlation remains a matter of debate. 

The correlation between mid and far IR luminosities of galaxies is more direct since both are produced by the re-emission of UV photons by dust, but the physics of the emitting sources is quite complex.  Above 60\,$\mu$m, the radiation of a galaxy is dominated by the grey body emission of ``big dust grains'' reaching an equilibrium temperature of 20-50 K. At shorter wavelengths, the spectrum of a galaxy results from the combination of :

- the continuum emission due to stochastically heated, warm dust grains, fluctuating around temperatures of a few hundred degrees. The carriers associated with this emission are commonly considered to be Very Small Grains \citep[VSG]{DBP90}.

- the PAH (polycyclic aromatic hydrocarbon) bands due to stochastically heated molecules producing broad features in emission at 3.3, 6.2, 7.7, 8.6, 11.3 and 12.7\,$\mu$m.

- the absorption features due to silicates which are centered at 9.7 and 18\,$\mu$m. 

Extragalactic surveys in the MIR, FIR, sub-millimeter and radio have all been used to trace star formation without being affected by dust extinction. However, with existing datasets and instrument sensitivities, the MIR is the most efficient wavelength to detect galaxies in the LIRG regime up to $z\sim$ 2. Sub-millimeter surveys are limited to the mJy level where LIRGs are only detected up to $z\sim$ 0.5 and with a bias towards objects with cooler dust temperatures for a fixed IR luminosity \citep{chapman05}.
Deep radio surveys provided a picture consistent with the one derived from the MIR but to a shallower depth \citep{Haarsma00,E02,Elbaz03}. As a result, it is crucial to determine whether MIR properties of distant infrared luminous galaxies are similar to local ones and assess if the MIR can be used as a tracer of star formation in the distant universe. This is the goal of the present paper.

We will first compare the $L(IR)$ derived independantly from three tracers, namely the radio continuum at 1.4 and/or 8.5 GHz, the 15\,$\mu$m flux density from ISOCAM \citep{Cesarsky96} onboard the Infrared Space Observatory \citep[ ISO]{Kessler96} and the 24\,$\mu$m flux density from MIPS \citep[Multiband Imaging Photometer for Spitzer][]{Rieke04} onboard the Spitzer Space Telescope \citep{Werner04}. The flux density and variation of the k-correction at the
two MIR wavelengths are used to constrain the relative contributions of the three components of MIR light : VSGs, PAHs and Silicate absorption. Using a similar strategy, \cite{Elbaz05} and \cite{Teplitz05} showed evidence that the MIR SED distant LIRGs did exhibit PAH features in emission. We will extend this work statistically by producing a direct comparison of the positions of local and distant galaxies in luminosity-luminosity diagrams at the same rest-frame wavelengths. This will allow us to test a possible variation of the relative roles played by the previous three ingredients of MIR SEDs which would have strong implications on the derivation of the CSFH.

While the derivation of L(IR) from a radio measurement depends only on the radio--far infrared (FIR) correlation and radio spectral index, which is well constrained for star forming galaxies, it is strongly dependant on the library of template SEDs that is chosen when using the MIR. Hence we will compare three commonly used SED libraries -- CE \citep{CE}, DH \citep{DH02} and LDP \citep{L04} -- which present some noticeable differences in terms of the relative contributions of the three major components of MIR light.

We adopted a $H_{0}$=75~km~s$^{-1}$~Mpc$^{-1}$, $\Omega_{M}=0.3$, $\Omega_{\Lambda}=0.7$ cosmology throughout this paper.
\section{Sample}
The sample consists of 49 IR luminous galaxies detected with ISOCAM and MIPS, including a subsample of 17 radio-detected galaxies, located in the Great Observatories Origins Deep Survey in the Northern hemisphere  (GOODS-N) centered on the Hubble Deep Field North (HDFN). GOODS is a NASA Legacy Survey with Spitzer (Dickinson et al., in prep.) and with the Hubble Space Telescope (HST) Advanced Camera for Surveys (ACS) \citep{Giavalisco04}. 

The size of the sample is limited by the depth of the ISOCAM catalog at 15\,$\mu$m \citep{Aussel99}.
This catalog is complete at the 90\,\% level down to 100\,$\mu$Jy but  a total of 95 sources are detected down to $\sim$10\,$\mu$Jy among which 81\,\% have a spectroscopic redshift \citep{Cohen00,Wirth04}. We used a revised version of the \cite{Aussel99} catalog (H.Aussel, private communication): 15\,$\mu$m sources were attributed an optical counterparts from HST images and when two or more ISOCAM galaxies were closer than 12\arcsec, the optical position of the HST sources was used to deblend the MIR sources. We rejected ISOCAM sources for which two potential optical counterparts were closer than the FWHM of the ISOCAM PSF \citep[4.6\arcsec, ][]{okumura98} and kept only the sources with negligible residuals after deconvolution using the optical centers in the final catalog.  After extracting those, the catalog consists of 54 galaxies, all of which turned out to possess a MIPS counterpart. We checked all individual sources by eye in the MIPS-24 $\mu$m image and found that blending was not an issue for any of this sub-sample of galaxies. Finally, we removed the five active galactic nuclei (AGN) that were identified from their X-ray emission \citep{Fadda02,Alexander02} and the final sample consists of 49 galaxies.

The radio data used in this study came from VLA A-array observations  done by \cite{Richards00} and new additional 26 hours of  B-array data  acquired by Morrison et al. (2006). The combined data sets and new  reductions used here yielded an improved rms ~ of 5.3uJy at the phase  center of the reduced map and provided a significant fraction  of new  faint radio detections (Morrison et al. 2006). The final sample is  described in Tables~\ref{TAB:sample1},~\ref{TAB:sample2}.

The astrometry of the MIPS 24\,$\mu$m images is good to ~0.2$\arcsec$ based on
a cross correlation with the IRAC images. Fewer than 10 sources, none of
which are relevant for this paper, are extended in the MIPS images which
have a spatial resolution of 5.7$\arcsec$ (PSF FWHM). Furthermore, almost all
24\,$\mu$m sources (down to $\sim$20\,$\mu$Jy, 5-$\sigma$) are found to have an IRAC 3.6\,$\mu$m counterpart.
24\,$\mu$m fluxes were determined by fitting a 24\,$\mu$m PSF at the
position of the IRAC sources, using a technique similar to DAOPHOT.
The technique will be described in greater detail in the GOODS MIPS
catalog paper (Chary et al. 2005, in prep.). This technique alleviates the effect
of confusion noise and provides a direct association between the 24\,$\mu$m 
source and its shorter wavelength counterpart.

The sample of local galaxies which will be compared to the distant luminous infrared galaxies consist of 154 galaxies (z$\leq$0.1) with flux densities measured at 12, 25, 60 and 100\,$\mu$m with IRAS and at 7 and 15\,$\mu$m with ISOCAM as described in \cite{CE} and \cite{E02}. Only a subsample of these objects have 
1.4 GHz radio flux densities.

\section{Total IR luminosity and SED libraries}

\subsection{SED libraries and decomposition between VSGs and PAHs}

\begin{figure}
       \resizebox{\hsize}{!}{\includegraphics{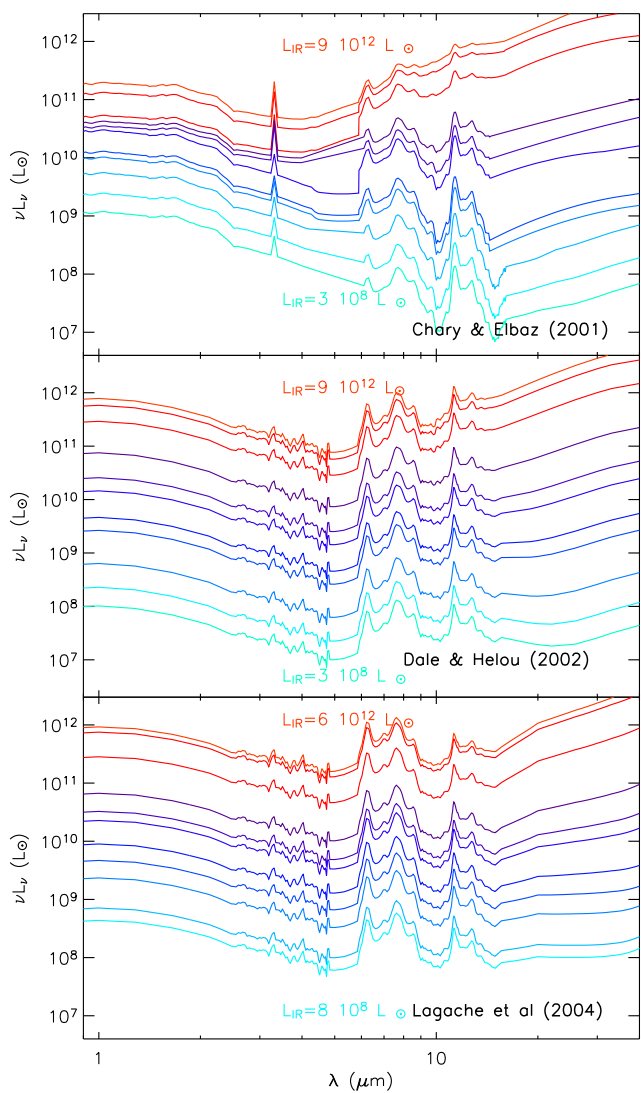}}
      \caption { 
{Evolution of the SED templates with infrared luminosity in the 3-30 $\mu$m range 
\bf{top :}} Chary \& Elbaz SED templates (2001) 
{\bf{center :}} Dale \& Helou SED templates (2002)
{\bf{bottom :}} Lagache et al (2004)
}
\label{FIG:SEDs}
\end{figure}

\begin{figure}
 \resizebox{\hsize}{!}{\includegraphics{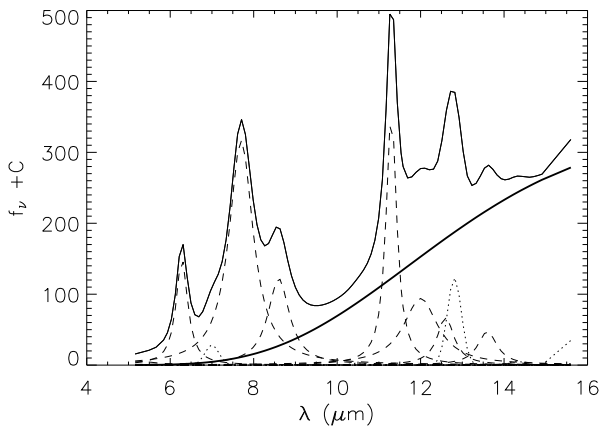}}
\caption{Demonstration of the decomposition method on the template galaxy with
     $L(IR) = 9\times10^{11}\;L_\odot$, in the DH library.
     The spectrum (thin line) is fitted with the linear combination of
     a hot grain continuum (bold line), PAH bands (dashed lines) and ionic
     lines (dotted lines).}
\label{fig:decomp}
\end{figure}

\begin{figure*}
  \resizebox{\hsize}{!}{\includegraphics{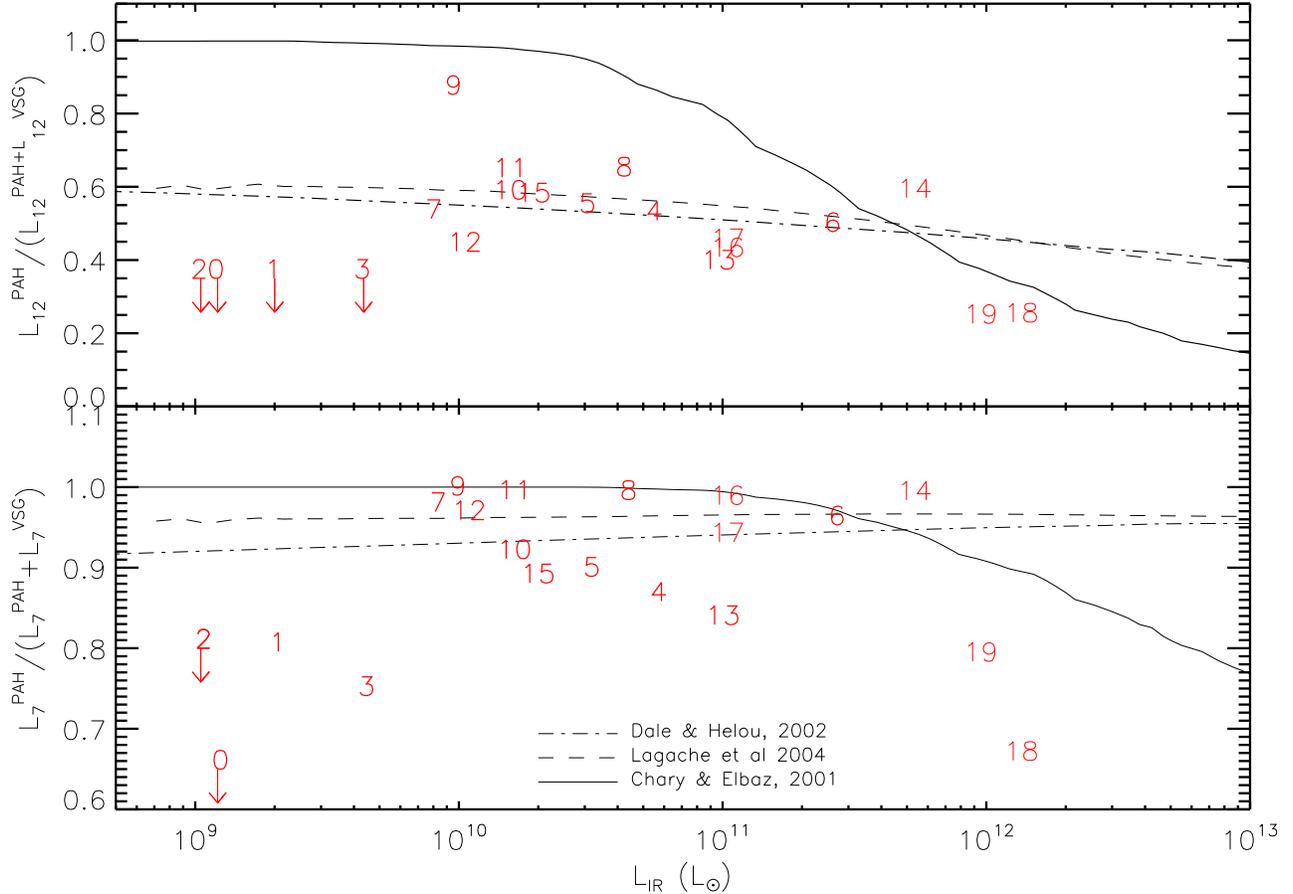}}
\caption{Evolution of the $L_{PAH}$/($L_{PAH}$+$L_{VSG}$) with L(IR) at
7$\mu$m (up) and 12 $\mu$m (down). Sample of galaxies :
{\bf{0:}} SBS 0335 (0.03 Z$_{\odot}$)
{\bf{1:}} IIZw 40 (0.2 Z$_{\odot}$),
{\bf{2:}} NGC 5253 (0.2 Z$_{\odot}$),
{\bf{3 :}} NGC 1140 (0.4 Z$_{\odot}$),
{\bf{4 :}} NGC 520,
{\bf{5 :}} M 51,
{\bf{6 :}} NGC 3256,
{\bf{7 :}} Cen A,
{\bf{8 :}} M 82,
{\bf{9 :}} IC 342 (central starforming region),
{\bf{10 :}} NGC 6946,
{\bf{11 :}} NGC 253,
{\bf{12 :}} Circinus,
{\bf{13 :}} NGC 1365,
{\bf{14 :}} NGC 6240,
{\bf{15 :}} M 83,
{\bf{16 :}} NGC 4038/39 \citep[ overlapping region of the Antennae galaxies]{Mirabel98},
{\bf{17 :}} nucleus of NGC 4038 \citep{Mirabel98},
{\bf{18 :}} Arp 220 \citep{charmandaris99},
{\bf{19 :}} IRAS 23128--5919 \citep{charmandaris2002}.
Galaxies number 4 to 19 have metallicities between 0.6 and 2 $Z_\odot$ 
while galaxies number 0 to 3 have metallicity $\leq$ 0.4 $Z_\odot$ as indicated within parenthesis.
}
\label{fig:decomp1}
\end{figure*}

We briefly summarize in the following the origin of the three libraries of template SEDs -- CE, DH, LDP -- (see Fig.~\ref{FIG:SEDs})  and we refer the reader to the associated papers for more details. While the CE templates were built to reproduce the observed local luminosity-luminosity correlations from 7 to 850\,$\mu$m, the DH ones were designed to fit the observed color-color relationships between local galaxies. The DH library consists of 64 templates sorted as a function of the $f_{\nu}^{60}$/$f_{\nu}^{100}$ ratio in the IRAS FIR bands. In order to compare this library to the CE and LDP ones, we attributed a total IR luminosity, $L(IR)$, to each SED using Eq.~\ref{EQ:f60f100} that we derived from the fit of the $log_{10}[f_{\nu}^{60}$/$f_{\nu}^{100}]$ versus $log_{10}[L(IR)]$ relationship \citep[see Fig.6a from][]{BGS} for objects in the Bright Galaxy Sample. The coefficients and dispersion that we used are given in Eq.~\ref{EQ:f60f100}. The resulting SEDs exhibit total IR luminosities in the range $L(IR)$= 2.1$\times$10$^8$ -- 2.2$\times$10$^{14}$ L$_{\odot}$ (see Fig.~\ref{FIG:SEDs}) and fit the local luminosity-luminosity correlations used by CE with a comparable quality.
\begin{equation}
log_{10}\left[f_{\nu}^{60}/f_{\nu}^{100} \right] = 0.128 \times log_{10}[L(IR)] - 1.611_{-0.119}^{+0.112}
\label{EQ:f60f100}
\end{equation} 
The shape of the 46 LDP starburst templates ($L(IR)$= 3$\times10^9$ -- 6$\times10^{12}$ L$_{\odot}$) in the MIR present the smallest evolution with $L(IR)$. 
They result from a revision by LDP of the \cite{L03} templates where 
the continuum in the 12-30 $\mu$m has been modified to allow a better fit of the MIPS-24\,$\mu$m number counts. 
Another 46 normal templates were also built by \cite{L03} to reproduce the population of spiral galaxies which radiate more than half of their energy in the optical. Since infrared luminous galaxies do not fall in this category, we did not use these SEDs for the estimation of L(LIR,MIR). However, we will present their location in the MIR color-color diagrams of Sect.~\ref{SEC:mirsed}.

We have used the method of \citet{verstraete01} and \citet{galliano+05} to decompose 
each SED from all three libraries into the sum of broad emission lines at 5.3, 5.7, 6.2, 7.7, 8.6, 11.3, 12.0, 12.7, 13.6 and 14.3\,$\mu$m due to PAHs and warm dust continuum due to VSGs, hereafter called the PAH and VSG contributions. 
Lorentzian profiles were adopted for the emission features
\citep[as proposed by][]{boulanger98,verstraete01}, with line widths and intensities as free parameters. The VSG continuum was
modeled with a modified black body, $\propto B_\nu(T)\times\nu^\beta$, $\beta=1$
being the emissivity; the temperature, $T$, and the intensity of this component
are free to vary, in order to minimise the $\chi^2$ of the fit.
The decomposition of the DH SED with $L(IR)$ = 9$\times$10$^{11}$ L$_{\odot}$ is presented in Fig.~\ref{fig:decomp}. 
This technique provides excellent fits for all SEDs regardless of luminosity and without introducing a silicate feature in absorption at 9.7\,$\mu$m. The presence of a trough around 10\,$\mu$m that is visible mostly in the CE SEDs below 5$\times$10$^{11}$ L$_{\odot}$ (Fig.~\ref{FIG:SEDs}) is perfectly reproduced by the depression between the right wing of the 8.6~$\mu m$ feature and the left wing of the 11.3~$\mu m$ one. It has been shown by various authors that the MIR SED of local star forming galaxies can be reproduced in the absence of silicate absorption at 9.7~$\mu m$ which is difficult to separate from PAH bands when the latter are intense \citep[e.g.][]{sturm+00}. PAHs are always strong in the three SED libraries both around 7\,$\mu$m and 12\,$\mu$m, except for the most luminous galaxies in the CE library, but this is also where no trough is noticeable (see Fig.~\ref{fig:decomp1}). 

This decomposition allows to follow the relative contribution of each component of the SEDs, i.e. PAHs and VSGs, as a function of wavelength for all SEDs and to study its evolution with L(IR). We computed this evolution with L(IR) in two passbands centered on the two strongest PAH features at 7 and 12\,$\mu$m. These two passbands correspond to the rest-frame central wavelenths probed by the ISOCAM-15\,$\mu$m and MIPS-24\,$\mu$m filters at $z=$ 1. The two components for each passband are  hereafter called $L_{7}^{\rm PAH}$, $L_{7}^{\rm VSG}$ and $L_{12}^{\rm PAH}$, $L_{12}^{\rm VSG}$. The DH and LDP libraries present a similar behavior (see Fig.~\ref{fig:decomp1}), i.e. very little variation of the contribution of PAHs to these bands over the whole range of total IR luminosities: PAHs are responsible for $\sim$90\,\% ($\sim$50\,\%) of the light in the 7\,$\mu$m (12\,$\mu$m) band. 

On the contrary, while the contribution of PAHs remains larger than 80\% in the 7 $\mu$m band for the CE library, it becomes smaller than that of the VSGs in the 12 $\mu$m band $L(IR)$ = 5$\times$10$^{11}$ L$_{\odot}$, and reaches ~20\% for the highest luminosities. A direct comparison with a sample of local galaxies for which both ISOCAM spectroscopy and IRAS broadband fluxes are available suggests that the behavior of the DH and LDP templates is more representative of the 7-12 $\mu$m color for local galaxies although we do see a trend for the contribution of PAHs to decrease in the ULIRG regime (see galaxies 18 and 19 in Fig.~\ref{fig:decomp1}). These local galaxies are comprised of local spirals \citep{roussel+01}, starbursts \citep{laurent+00} and dwarf galaxies \citep{madden+05} observed with the Circular Variable Filter of ISOCAM. Note that the LDP templates provide a ratio of the 12 over 7\,$\mu$m flux densities which does not vary with luminosity, contrary to what is observed for local galaxies, and is twice smaller than the
observed average ratio (see Sect.~\ref{SEC:mirsed}).

In Fig.~\ref{fig:decomp1}, galaxies number 4 to 19 have metallicities between 0.6 and 2 $Z_\odot$, typical of relatively massive galaxies at $0 < z < 1$.  Four galaxies (numbers 0 to 3 in Fig.~\ref{fig:decomp1}), have metallicity lower than  0.4 $Z_\odot$.  Their properties may be reflective of primeval galaxies,
for which local SED libraries might not be valid in the derivation of total IR luminosities (see \cite{galliano+03,galliano+05,engelbracht+05}.

\subsection{Derivation of the total infrared luminosity (8--1000\,$\mu$m)}
The correlations between the FIR (40--120\,$\mu$m, $L_{\rm FIR}$) or total IR (8--1000\,$\mu$m, $L(IR)$) luminosity with the radio and MIR can be used to derive independent estimates of $L(IR)$. We use $L(IR,radio)$, L(IR,15) and $L(IR,24)$ for the infrared luminosity derived from the radio, ISOCAM-15 and MIPS-24 passbands respectively. $L_{\rm FIR}$ and $L(IR)$ are related by Eq.~\ref{EQ:FIR} computed from the Bright Galaxy Sample \citep{BGS} where we use the definitions of \cite{Helou88} for $L_{\rm FIR}$ and \cite{SM} for $L(IR)$.

\begin{equation}
L_{IR}=1.91 (\pm 0.17) \times L(FIR).
\label{EQ:FIR}
\end{equation}

For the radio-FIR correlation, we used a q-parameter of 2.34 as defined in Eq.~\ref{EQ:radio} \citep{Condon92,Yun01}:

\begin{equation}
q=log_{10}\left( 
\frac{L(FIR)({\rm W})}{3.75 \times 10^{12}({\rm Hz})} 
\frac{1}{L_{1.4 {\rm GHz}}({\rm WHz} ^{-1})}
\right).
\label{EQ:radio}
\end{equation}
The rest-frame 1.4 GHz monochromatic luminosity was derived from the observed 1.4 or 8.5 GHz flux densities assuming a radio spectral slope for star forming galaxies of $\alpha= 0.8 \pm0.15$ , \citep[$S_{\nu} \propto \nu ^ {-\alpha }$]{Yun01}. 

To derive $L(IR)$ from the MIR (15 and /or 24 $\mu$m) , we used the strategy advocated in \cite{CE}. Each one of the template SEDs is redshifted to the distance of the observed source. Then, the SED whose MIR flux density is the closest to the observed one is used to derive $L(IR)$ after it is normalized to the exact observed flux density. 

\section{Global consistency of the 15$\mu$m, 24$\mu$m and radio star formation indicators}
\label{SEC:radio}
\begin{figure*}
   \resizebox{\hsize}{!}{\includegraphics{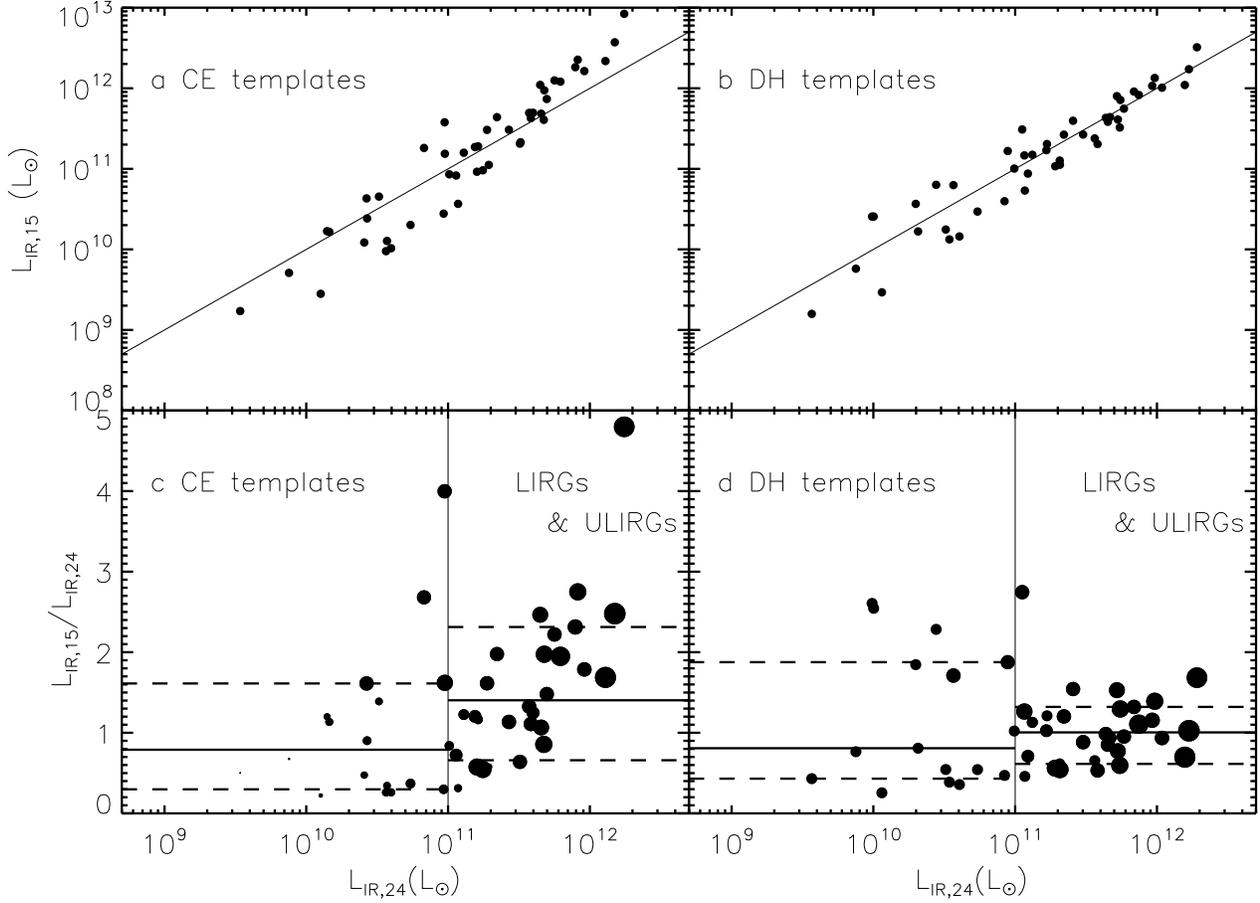}}
\caption{ 
{\bf{a :}} $LIR_{15}$ as a function of  $LIR_{24}$ for the SED of CE;
{\bf{b :}} Evolution of the ratio $LIR_{15}$/$LIR_{24}$ with respect to $LIR_{24}$ for the SED of DH;
 {\bf{c :}} $LIR_{15}$ as a function of  $LIR_{24}$ for the SED of CE;
{\bf{d :}} Evolution of the ratio $LIR_{15}$/$LIR_{24}$ with respect to $LIR_{24}$ for the SED of DH;
the size of the circles is proportionnal to z. 
}
\label{FIG:1524}
\end{figure*}
\begin{table*}
\centering
\begin{tabular}{lcccc}
\hline
\hline
                                       &                        & \multicolumn{3}{c}{$L(IR,15)$/$L(IR,24)$}    \\    
              & z                      &CE                       & DH        & LDP\\    
\hline
all galaxies                            &$0.84_{-0.38}^{+0.17}$  &$1.2_{-0.7}^{+0.8}$  &$1.0_{-0.4}^{+0.8}$ & 0.5$^{+0.4}_{-0.2}$  \\
 L(IR,24)$<10^{11}$ L$_{\odot}$         &$0.47_{-0.22}^{+0.09} $ &$0.8_{-0.4}^{+0.8}$  &$0.8_{-0.4}^{+1.0}$ & 0.8$^{+0.9}_{-0.5}$  \\
 L(IR,24)$\geq 10^{11}$ L$_{\odot}$     &$0.85_{-0.10}^{+0.19} $ &$1.3_{-0.6}^{+1.0}$  &$1.0_{-0.4}^{+0.4}$& 0.5$^{+0.2}_{-0.2}$   \\
\hline
\hline
\end{tabular}
\label{TAB:1524}
\caption{Median and 68\,\% dispersion (around the median) of the ratio $L(IR,15)$/$L(IR,24)$ estimated with the SED of CE, DH and LDP.}
\end{table*} 
\begin{figure*}
  \resizebox{\hsize}{!}{\includegraphics{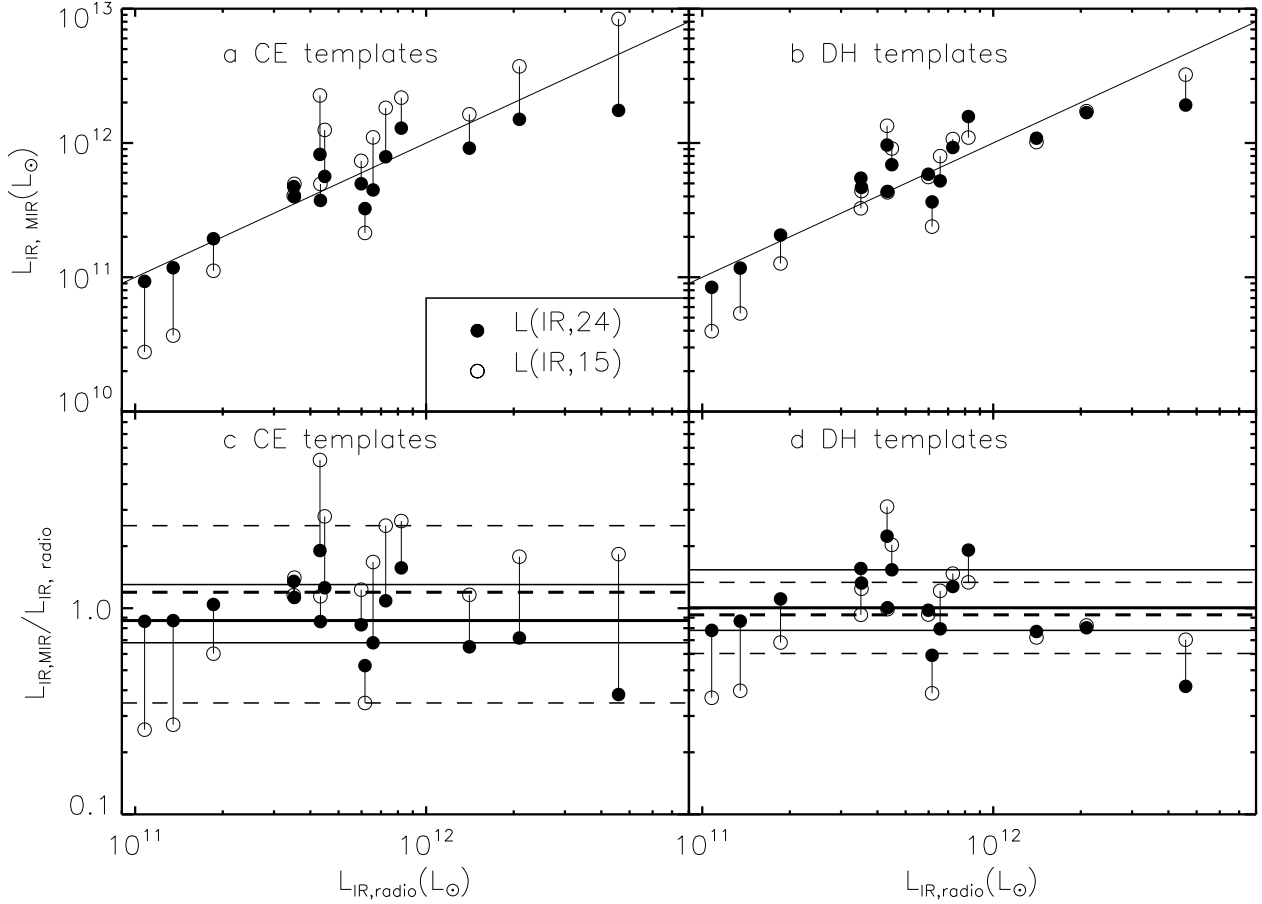}}
\caption{ Comparison of the total IR luminosity (8--1000\,$\mu$m) derived from the radio (from the 1.4 and/or 8.5 GHz bands) and the MIR at 15\,$\mu$m (open circles) and 24\,$\mu$m (filled circles). Figs a,c and Figs b,d were produced using the CE and DH libraries of template SEDs respectively, to derive $L_{\rm IR, MIR}$. The solid line in Figs.a,b is the one-to-correlation between the two L(IR).
The solid and dashed lines in Figs c,d represent the median (thick lines) and 1-$\sigma$ (68\,\%) dispersion (thin line) around this median for the 24 (solid) and 15\,$\mu$m (dashed) data respectively.
}
\label{FIG:rad_IR}
\end{figure*}
\begin{table*}[t]
\centering
\begin{tabular}{cccccccc}
\hline
                & ~~~z &\multicolumn{6}{c}{$L(IR,MIR)$/$L(IR,radio)$}  \\    
        &  & \multicolumn{3}{c}{15\,$\mu$m}& \multicolumn{3}{c}{24\,$\mu$m}\\
                                          &                          & CE &DH & LDP   & CE &DH & LDP   \\    
\hline
&&&&&&&\\ 
  $10^{11} \leq $L(IR) $\leq 10^{12}$ L$_{\odot}$  & $0.85~_{-0.4}^{+0.4}$ &$1.2~_{-0.8}^{+1.4}$ &$1.0~_{-0.5}^{+0.6}$&0.6$^{+0.3}_{-0.3}$&$1.0~_{-0.2}^{+0.3}$ &  $1.1~_{-0.3}^{+0.4}$ & 1.3$^{+0.8}_{-0.5}$ \\
&&&&&&&\\ 
\hline
\end{tabular}
\caption{Median and 68\,\% dispersion around the median of the ratio $L(IR,MIR)$/$L(IR,radio)$ estimated with the template SEDs of CE, DH and LDP.}
\label{TAB:LIR}
\end{table*} 
The 15 and 24\,$\mu$m passbands probe different sources of MIR emission at the median redshift ($z\sim$ 0.85) of the sample considered here. At this distance, the 15\,$\mu$m band samples the PAH emission centered around 7.7\,$\mu$m while the 24\,$\mu$m band measures the combined emission from VSGs and the 11.3 and 12.7\,$\mu$m bands. However, both bandpasses provide consistent values for L(IR) (see Fig.~\ref{FIG:1524}) with a median ratio of 1 and an rms of 40\,\% when using the DH library (LIRGs/ULIRGs only, i.e. third line of  Table~1). This dispersion is the same as the MIR versus L(IR) one for local galaxies which suggests that the MIR SED of galaxies did not strongly evolve with redshift up to $z\sim$ 1. 

The 18 galaxies of the sample for which both MIR and radio data are available are located at redshifts between $0.3 < z < 1.3$. The total IR luminosity derived from 24\,$\mu$m and the radio are consistent within 40\,\% (rms defined as the 68\,\% confidence level, see Fig.~\ref{FIG:rad_IR}), in agreement with previous studies \citep{E02,Garrett02,Gruppioni03,Appleton04}. The median ratio of L(IR,MIR) over L(IR,radio) is $\sim$ 1 (see Table~\ref{TAB:LIR}) but the dispersion around this value depends on the MIR wavelength that is used and on the library of template SEDs. Note that the 40\,\% dispersion of L(IR,MIR)/L(IR,radio) is comparable to the 46\,\% dispersion for L(FIR)/L(IR,radio) found for the 162 local IR luminous galaxies of \cite{Yun01} and \cite{Reddy04}. Hence, the agreement between MIR and radio derived total IR luminosities at $z\sim$ 1 is at least as good as in the local universe, which suggests that the SED of galaxies in the IR were not strongly different at $z\sim$ 1 from what it is today.

The 24\,$\mu$m passband gives the lowest dispersion, with an rms of $\sim$ 30 and 40\,\% for the CE and DH libraries respectively. At 15\,$\mu$m, the dispersion becomes as large as 100\,\% for the CE library and 60\,\% for DH. The very good agreement of L(IR,24) with L(IR,radio) supports the idea that the radio remains a robust SFR indicator up to $z\sim$ 1.3. This in turns implies that the 15\,$\mu$m passband is a less robust tracer of L(IR) than 24\,$\mu$m. Since the median redshift of those galaxies is $z\sim$ 0.85, this suggests that the warm dust continuum is a better tracer of L(IR) than PAHs. 

A comparison between the results obtained with the CE and DH libraries shows that systematic effects are observed for L(IR,24) $\geq$ $10^{11}$ L$_{\odot}$.
We find that the L(IR,15) derived from the CE library is 30\,\% larger than L(IR,24) and that the ratio of both luminosities presents a dispersion between 60 and 90\,\% (see Table~1). The major difference between the CE and DH libraries is the progressive dilution of PAHs by the VSG continuum as a function of increasing L(IR) for the CE SEDs (see Fig.~\ref{FIG:SEDs}) If this dilution were to be excessive or if PAHs were to be stronger than predicted by this library then one would expect the CE SEDs to overestimate the L(IR) derived from a flux density measured around the peak of PAHs emission around 8\,$\mu$m. This would explain the excess ratio of L(IR,15) over L(IR,24). And indeed, when compared to an independant tracer of L(IR), the radio, we find that L(IR,15)/L(IR,radio) is 20\,\% larger than one, with a dispersion larger than with the DH library (see Table~\ref{TAB:LIR}).
When computed using the LDP library, L(IR,24) is twice larger than L(IR,15) (see Table~1). The origin of this discrepancy is opposite to the one with CE, i.e. the strength of PAHs around $\lambda\sim$ 8\,$\mu$m (in the rest-frame of the $z\sim$ 0.85 galaxies) is too strong for these luminosities. Indeed the L(IR,MIR) predicted with the LDP library exhibit a systematic offset with respect to L(IR,radio) (see Table~\ref{TAB:LIR}). To avoid confusion, we did not plot those points in Fig.~\ref{FIG:rad_IR}. This discrepancy results from the 7.7\,$\mu$m PAH feature being too strong and the continuum plus PAH emission around 12\,$\mu$m being too faint in their SED library. Note that LDP designed their library
in order to better reproduce the 24\,$\mu$m galaxy counts observed with Spitzer. 

Finally, we note that the three galaxies with redshifts close to $z\sim$ 0.5 -- for which {\bf{$L(IR,radio)\leq$ 3$\times$10$^{11}$ L$_{\odot}$}} -- exhibit a $L(IR,15)/L(IR,radio)$ ratio lower than 1 both for CE and DH (Fig.~\ref{FIG:rad_IR}). At $z\sim$ 0.5, the 15\,$\mu$m passband probes the 10\,$\mu$m trough in the rest-frame which might be underestimated by those templates. The number of galaxies at this redshift for which both MIR and radio data are available is too small in the present sample to derive
any strong conclusion. However, we will address this point by comparing the MIR sample to local galaxies in the next section.

\section{Testing the redshift dependence of mid-infrared SEDs}
\label{SEC:mirsed}
\begin{figure}
       \resizebox{\hsize}{!}{\includegraphics{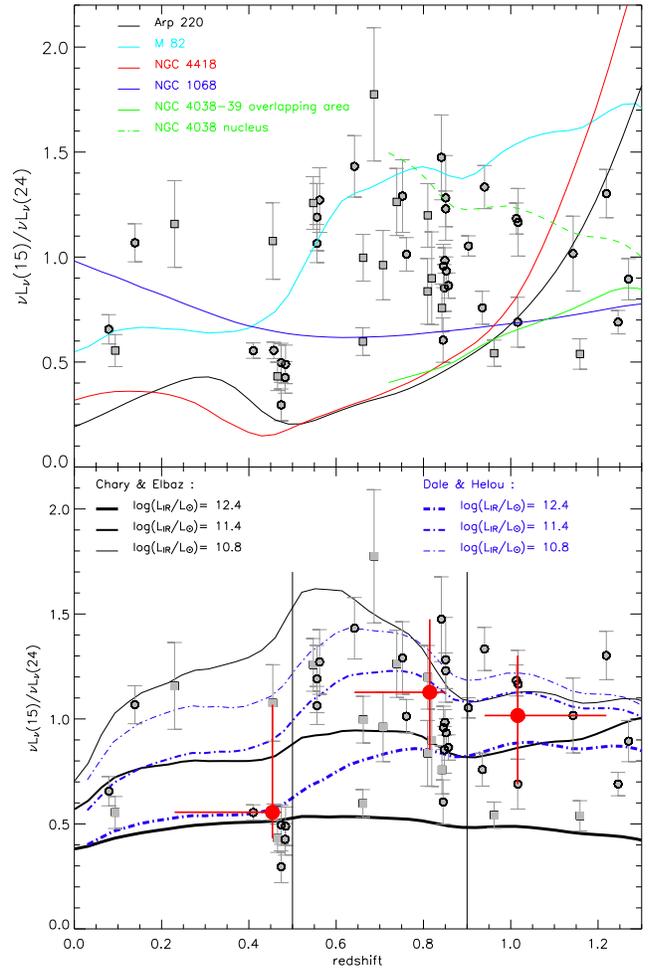}}
      \caption { $\nu L_{\nu}^{15 \mu m}/\nu L_{\nu}^{24 \mu m}$ as a function of the redshift. {\bf{Black and grey circles :}} our distant sample of galaxies. {\bf{Black and grey squares : }} data presented in \citet{Elbaz05}.{\bf{ Red circles :}} : median $\nu L_{\nu}^{15 \mu m}/\nu L_{\nu}^{24 \mu m}$ in each redshift bin with 1$\sigma$ error bars.} 
      \label{FIG:15-241}
\end{figure}
In order to check whether distant and local luminous IR galaxies exhibit consistent MIR properties, we have compared the evolution of the observed
24 $\mu$m over 15 $\mu$m luminosity ratio with redshift to prototypical starbursts and active galactic nuclei (fig.~\ref{FIG:15-241}a) and to SEDs such as CE and DH (fig~\ref{FIG:15-241}b). 
We have also selected two redshift bins ($z\sim$ 0.6 and 1) such that the 15 and 24\,$\mu$m passbands probe rest-frame wavelengths for which data on local galaxies are already available. Local and distant galaxies are represented with filled triangles and circles respectively in the Figs.~\ref{FIG:12/7-1},~\ref{FIG:10-15-1}. We distinguish three domains in luminosity for ``normal" galaxies, ``LIRGs" and ``ULIRGS", corresponding to [10$^{10}$-10$^{11}$], [ 10$^{11}$-10$^{12}$] and greater than 10$^{12}$ L$_{\odot}$ range. In both figures, the median ratio of the luminosities for local and distant galaxies are marked with dark solid triangles and circles respectively. The 1$\sigma$ error bars were computed by the quadratic sum of a poissonian error and an error computed to include 68\,\% of the galaxies in each bin. We will center the discussion on the normal and LIRG populations due to a lack of objects outside of these domains. 
We will start by presenting the redshift evolution of the observed 15 over 24\,$\mu$m color as a test for the presence of PAHs in distant galaxies.

\subsection{Testing the presence of PAHs in distant galaxies using the 15 over 24\,$\mu$m ratio}

In Fig.~\ref{FIG:15-241}a, we compare the location of the GOODS-N galaxies (circles) in the 24 over 15\,$\mu$m luminosity ratio versus redshift space to several prototypical local SEDs. This figure reproduces the Fig.3 of \cite{Elbaz05} from which the square points are taken.
The closest starburst M82 ($L(IR)=4\times10^{10}$ L$_{\odot}$), the merging system of the Antennae (NGC 4038-39; LIRG,  $L(IR)=1.4\times10^{11}$ L$_{\odot}$, \cite{Mirabel98}), the highly obscured LIRGs NGC 4418 \citep{Spoon01}, the classic, but atypical, ULIRG, Arp 220 \citep{charmandaris99} and the Compton thick Seyfert II NGC 1068 \citep{LeFloch01} are compared to our distant LIRGs. 

The templates can be divided into three families: 
\begin{enumerate}
\item {\it Starbursts}. M 82 and the overlapping region of the Antennae galaxies \cite{Mirabel98,Vigroux96}: 
the SEDs of M82 and of the overlapping region of the Antennae  galaxies (Mirabel et al. 1998, Vigroux et al. 1996) globally  reproduce the behavior of the observed data but tend to respectively  over- and under-estimate the L15/L24 ratio above z=0.5.
This could be explained either by the warm dust continuum being too faint or the silicate absorption feature being too weak.

\item {\it Heavily obscured starbursts}: Arp 220, NGC 4418 and the nucleus of NGC 4038.

NGC 4418 is either a strongly embedded AGN or a compact starburst (Evans et al, 2003). The MIR spectrum shows strong silicate absorption without strong evidence for PAH features ( Spoon et al, 2001). 

The MIR spectrum of Arp 220 shows characteristics of a highly obscured continuum with silicate absorption (like NGC4418) and a starburst component characterized by PAH emission (like M82) (Spoon et al, 2004). ²

All three cases predict a low value for $L_{15}$/$L_{24}$ at z$\leq$0.9  and a high value at z $\geq$1.1. This is probably due to a combination of strong silicate absorption and continuum emission from warm dust.

\item {\it Active galactic nucleus (AGN)}. NGC 1068 harbors a central black hole with a circumnuclear starburst. 75 \% of the MIR emission is dominated by the AGN emission (Le Floc'h et al, 2001). The MIR spectrum is dominated by a strong continuum, without PAH. Its $L_{15}$/$L_{24}$ ratio is too weak for 68 \% of the sample due to its flat SED and the absence of PAH features in emission.

\end{enumerate}

In Fig.~\ref{FIG:15-241}b, we have indicated three bins of redshift (0$-$0.5, 0.5$-$0.9, 0.9$-$1.3) which correspond approximately to three bins in luminosity because of selection effects: [$3 \times 10^8-3 \times 10^{11}$], [$3 \times 10^9-7 \times 10^{11}$] and [$10^{10}-3\times 10^{12}$] L$_{\odot}$. In each bin, the red dot is associated to the median value of the data with an error bar including 68\,\% of the sample.  The position of the red dots confirms the conclusion of Elbaz et al. (2005) that the  $L_{15}/L_{24}$ ratio increases from $z\sim$ 0.2 to 0.7 due to the entrance of the 7\,$\mu$m PAH emission feature into the ISOCAM-15\,$\mu$m band.  

The template SEDs from CE and DH are represented for three luminosities, $L(IR)=10^{10.8}$ (thin line) , $10^{11.4}$ (thick line) and $10^{12.4}$ (very thick line) L$_{\odot}$. The global behaviour of the data is well represented by both families of synthetic templates. However, for the CE templates, the $z=$ 0.5$-$0.9 bin is better represented by SEDs less luminous than LIRGs while the observed galaxies are predominantly LIRGs. This confirms the previous statement that in the CE templates, the PAHs emission in IR luminous galaxies is underestimated by about 15-20\%, while this is not the case of the DH templates.

\subsection{$L_{12}$ versus $L_7$}
\label{SEC:L127}

\begin{figure*}
        \resizebox{\hsize}{!}{\includegraphics{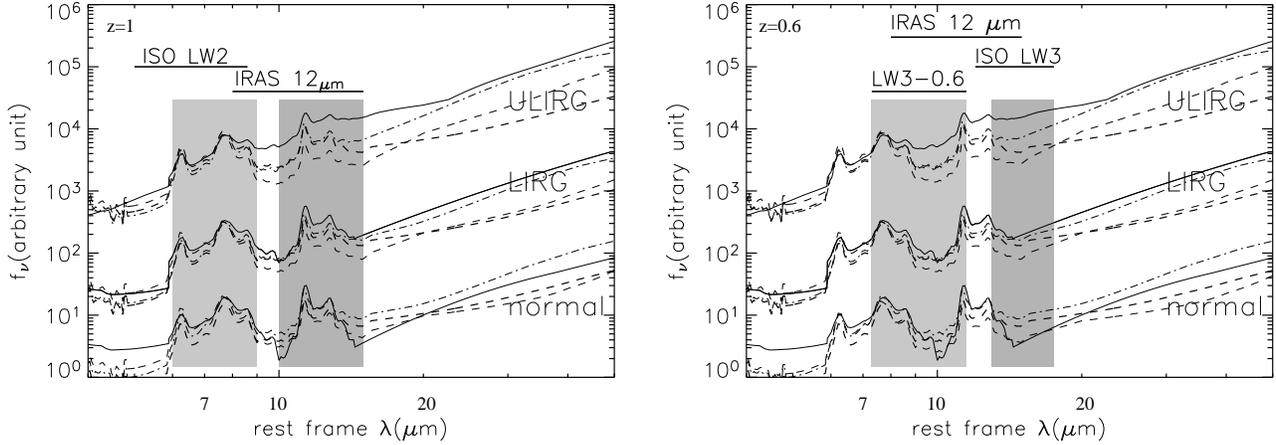}}
\caption{Relative positions of the ISOCAM-15\,$\mu$m and MIPS-24\,$\mu$m bands for galaxies located at {\bf{$z\sim$ 1 (a) and 0.6 (b)}} with respect to the observed ISOCAM-LW2, LW3 and IRAS-12\,$\mu$m bands. represented in the plots is $f_{\nu}$ in arbitrary units as a function of $\lambda$. Origin of the template SEDs at 3 different luminosities ($L(IR)=10^{10}$ L$_{\odot}$, "normal", $10^{11}$ L$_{\odot}$, "LIRG" and $10^{12}$ L$_{\odot}$, "ULIRG"): CE ({\bf{Solid line}}), DH ({\bf{Dash-dotted line}}), LDP ({\bf{Dashed line:}} Starburst template, {\bf{thick dashed line}} "cold" template).
\label{FIG:12/7-fil}
}
\end{figure*}

\begin{figure}
       \resizebox{\hsize}{!}{\includegraphics{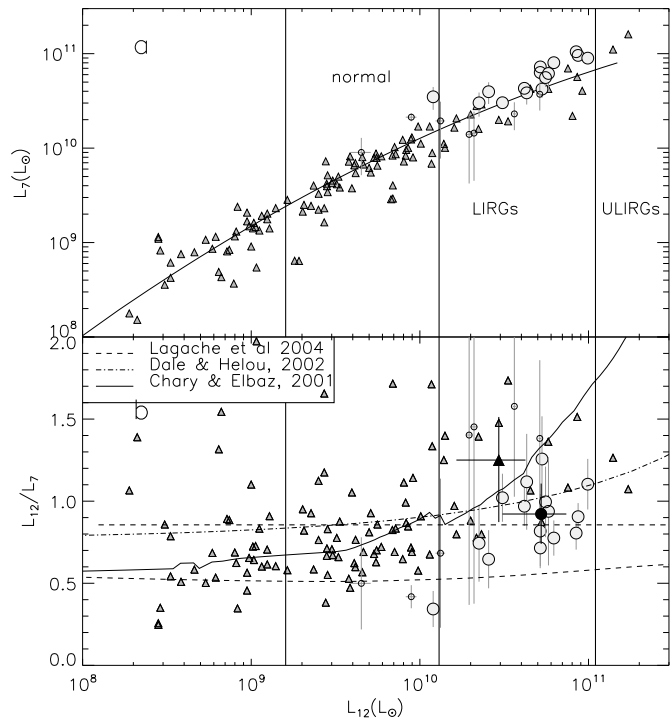}}
 \caption{Comparison between the rest-frame 7 and 12\,$\mu$m luminosities of local (filled triangles) and $z\sim$ 1 ( 0.8 $\leq$ z $\leq$ 1.25 ,{\bf{big filled circles :}} f15 $\ge$ 150 $\mu$Jy, {\bf{little filled circles :}} f15 $\le$ 150 $\mu$Jy) galaxies. $L_7$ and $L_{12}$ correspond to $\nu L_{\nu}$ at these central wavelengths for the local- ISOCAM-6.75\,$\mu$m and IRAS-12\,$\mu$m passbands - and $z\sim$ 1 - rest-frame bands corresponding to the observed ISOCAM-6.75\,$\mu$m and MIPS-24\,$\mu$m - galaxies. Two domains of luminosities are defined by vertical lines (see text) where the local and distant galaxies are compared. 
({\bf{a}}$) L_7$ as a function of $L_{12}$. The plain line is a polynomial of degree 2 fit to the local sample alone.
({\bf{b}}) $L_{12}$/$L_7$ ratio as a function of $L_{12}$. The median ratio for local and distant galaxies are indicated with dark triangles and circles respectively, in each luminosity bin. The lines indicate the prediction for the CE ({\bf{plain line}}), DH ({\bf{dash dot line}}), LDP cold ({\bf{thick dashed line}}) and starburst ({\bf{dashed line}}) templates.}
\label{FIG:12/7-1}
\end{figure}
\begin{figure}
       \resizebox{\hsize}{!}{\includegraphics{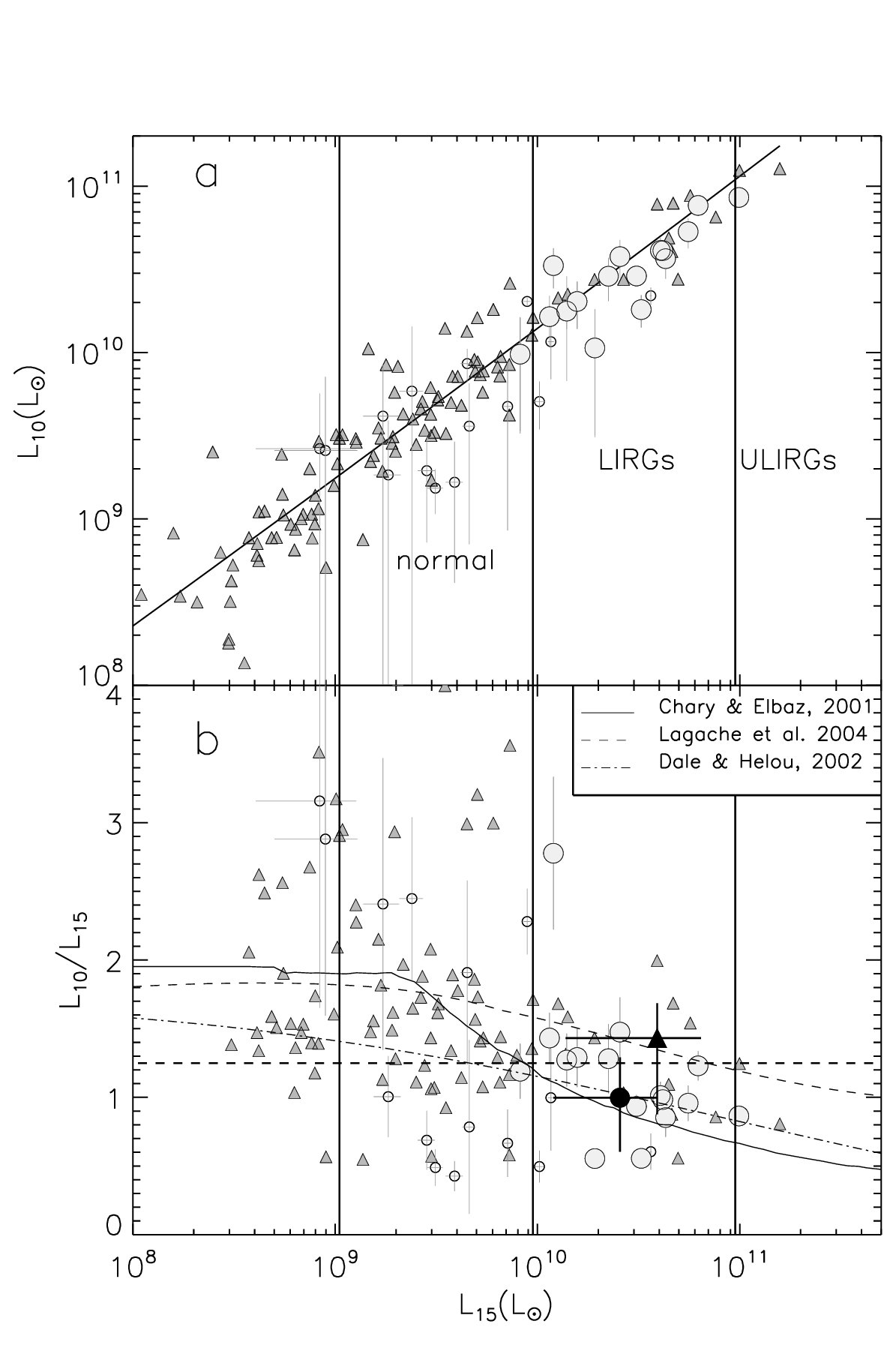}}
      \caption { 
Comparison of the rest-frame 10 and 15\,$\mu$m luminosities of local (filled triangles) and $z\sim$ 0.6 (0.4 $\leq$ z $\leq$ 0.8, {\bf{big filled circles :}} f15 $\ge$ 150 $\mu$Jy, {\bf{little filled circles :}}  f15 $\le$ 150 $\mu$Jy) galaxies. $L_{15}$ corresponds to $\nu L_{\nu}$ at this central wavelength for the ISOCAM-15\,$\mu$m and MIPS-24\,$\mu$m passbands for local and $z\sim$ 0.6 galaxies respectively. $L_{10}$  corresponds to the rest-frame emission measured in the ISOCAM-15\,$\mu$m band at $z\sim$ 0.6 and was derived from a combination of the IRAS-12\,$\mu$m and ISOCAM-15\,$\mu$m bands for local galaxies (see Eq.~\ref{EQ:relat}.
({\bf{a}}) $L_{10}$ as a function of $L_{15}$. The plain line is a polynomial of degree 2 fit to the local sample alone (here equivalent to a linear fit).
({\bf{b}}) $L_{10}$/$L_{15}$ as a function of $L_{15}$. The median ratio for local and distant galaxies are respectively indicated with dark triangles and circles in both luminosity bins. The lines indicate the prediction for the CE ({\bf{plain line}}), DH ({\bf{dash dot line}}), LDP cold ({\bf{thick dashed line}}) and starburst ({\bf{dashed line}}) templates.
} 

\label{FIG:10-15-1}
\end{figure}

At $z \sim 1$, the 15 and 24\,$\mu$m observations measure
rest-frame wavelengths of approximately 7.5 and 12\,$\mu$m,
respectively.  We will designate the rest-frame luminosities
at these wavelengths as $L_7$ and $L_{12}$.  $L_7$ measures
the ``PAH bump'' centered on the 7.7\,$\mu$m dominant PAH feature, while $L_{12}$ measures a combination of the 11.3 and 12.7\,$\mu$m PAH bands with the warm dust continuum
due to VSGs, as well as some contribution from the silicate
feature in absorption, if any.  We may compare $L_7$ and $L_{12}$ for the
high-redshift galaxies to data for local ($z \sim 0$)
galaxies observed with ISOCAM-LW2, centered
at 6.75\,$\mu$m, and IRAS 12\,$\mu$m
respectively.  Fig.~\ref{FIG:12/7-fil}a shows the
rest-frame wavelengths observed by the 15 and 24\,$\mu$m
bands at $z \sim 1$ as grey shaded regions, and compares them
to the ISOCAM-LW2 and IRAS 12$\mu$m bandpasses for local
galaxies. The comparison between local and distant galaxies
at these wavelengths is presented in Fig.~\ref{FIG:12/7-1}.
Since the redshifted ISOCAM and MIPS filters do not exactly match the observed ISOCAM and IRAS filters used for local galaxies, we computed a color correction for each individual galaxy, taking into account its proper redshift, using the SED libraries. This color correction factor is always smaller than 20\,\% and the choice of the CE or DH library makes less than a 5\,\% difference.

We extracted a subsample of 116 galaxies from the local sample, all of which have flux measurements at both 7 and 12\,$\mu$m. The distant sample was made of 26 distant galaxies with a flux at 15 and 24 $\mu$m and located in the 0.8-1.25 redshift range in the redshift range $0.8 < z < 1.25$,
with median redshift $z = 0.90$.

Distant galaxies follow the same trend as local galaxies in their $L_7$ versus $L_{12}$ correlation (Fig.~\ref{FIG:12/7-1}a). This shows that at least in the MIR domain, the SED of galaxies have not evolved strongly between $z\sim0$ and $z\sim$ 1. We do find an indication of marginal evolution: nearly all $z\sim$ 1 galaxies lie above the polynomial fit of degree 2 to the local sample (solid line in Fig.~\ref{FIG:12/7-1}a). We computed in Fig.~\ref{FIG:12/7-1}b the median $L_{12}/L_7$ ratio for local (filled black triangle) and distant (filled black circle) LIRGs. 
Distant galaxies exhibit a lower $L_{12}/L_7$ ratio than local ones 
in both luminosity bins (LIRGs- normal galaxies), with an apparently stronger difference in the lower luminosity bin.  However, this bin consists of only six galaxies, four of which are fainter than 75\,$\mu$Jy and therefore lie below the 60\,\% completeness regime of the ISOCAM image \citep{Aussel99}. The risk that the galaxies detected at this flux density level might not be representative of the average galaxy properties (Malmquist bias in the sense of favoring objects with strong 15\,$\mu$m flux densities and low $L_{12}/L_7$) is therefore not negligible. Hence we consider that our sample is not robust against biases outside of the LIRG regime.

In the LIRG luminosity bin, all but two of the galaxies fall above the 90\,\% completeness limit of 100\,$\mu$Jy. The median $L_{12}/L_7$ ratio for local galaxies lies above the 1-$\sigma$ dispersion (including error bars) around the median ratio for distant galaxies, i.e. two thirds of the distant galaxies present a $L_{12}/L_7$ ratio lower than the median value for local galaxies. Although, this effect needs to be consolidated by a larger sample, we believe that it is important to consider the possibility that we are already seeing some indication of evolution that will become stronger at larger redshifts. Possible explanations for this evolution are discussed in Sect.~\ref{SEC:discussion}.

The $L_{12}/L_7$ versus $L_{12}$ trend is globally fitted by the CE and DH libraries but not the LDP one, which systematically underestimates $L_{12}/L_7$ over all luminosities. This observation confirms the excess PAH emission, already discussed in Sect.~\ref{SEC:radio}, for this library.  Finally, we note that the CE library presents a rapid increase of $L_{12}/L_7$ with increasing luminosity in the IR luminous regime and that the L(IR) derived using this library might be overestimated by up to 50\,\% for distant ULIRGs. This can be explained by the same reasons discussed in the previous section, i.e. the weak emission from PAHs around 8\,$\mu$m in this library.

\subsection{$L_{10}$ versus $L_{15}$}
\label{sec:1015}
At $z \sim 0.6$, the 15 and 24\,$\mu$m observations
measure rest-frame wavelengths of approximately 9.4 and
15\,$\mu$m, respectively.  We will designate these rest-frame
luminosities as $L_{10}$ and $L_{15}$.  While both $L_{10}$
and $L_{15}$ contain emission from the VSG warm dust continuum,
only $L_{10}$ is affected by the silicate absorption feature.
Hence, the flux ratio of these two bands helps to quantify
the strength of the silicate absorption.  We may compare
$L_{15}$ for high-redshift galaxies to data for local
($z \sim 0$) galaxies observed with ISOCAM-LW3, centered
at 15$\mu$m.  There is no direct observation of $L_{10}$
for local galaxies, but we may derive it from a combination
of observations with ISOCAM-LW2 and IRAS-12$\mu$m as described below.  
We will designate this synthetic bandpass as ``LW3-0.6''.  We illustrate the local and
redshifted bandpasses for the $z \sim 0.6$ sample in
Fig.~\ref{FIG:12/7-fil}b.  The comparison between local and distant galaxies
at these wavelengths is presented in Fig.~\ref{FIG:10-15-1}.

We have compared 115 local galaxies that have measured 12 and 15\,$\mu$m flux densities to 35 distant galaxies at $z\sim$ 0.6 (0.4 $\leq$ z $\leq$ 0.8).  In order to probe the wavelength range associated with the 10$\mu$m feature, we constructed a synthetic passband centered at 10\,$\mu$m using a combination of the ISOCAM-LW3 (12--18\,$\mu$m) and IRAS-12\,$\mu$m (8--15\,$\mu$m) filters. The passband of the synthetic filter was chosen to reproduce the rest-frame wavelength range probed by the ISOCAM-LW3 filter for a galaxy located at $z\sim$ 0.6, i.e. centered at 9.4\,$\mu$m and therefore close to the 10 $\mu$m feature . We found that a combination of the ISOCAM-LW3 and IRAS-12\,$\mu$m filters could provide an estimate of the synthetic passband (hereafter called LW3-0.6) flux density with an accuracy of 20\,\%. 
We optimized this combination for CE and DH templates independantly but found that the difference between
both computations is only 10\,\% (see Eq.~\ref{EQ:relat} computed for galaxies which $L(IR) \geq 4.7\times10^{10}$ L$_{\odot}$). The LW3-0.6 filter is represented with a line in fig.~\ref{FIG:12/7-fil}.
At $z=$ 0.6, the MIPS-24\,$\mu$m passband is roughly equivalent to the rest-frame ISOCAM-LW3 filter; however, 
we calculated a filter correction to take into account the k-correction and the difference of filter shapes. 
The correction is estimated at 10\,\% when using the DH templates and 30 \% for CE.
\begin{equation}
\begin{split}
&L_{LW3-0.6}^{CE} = 1.64_{-0.42}^{+0.14} \times L_{12}       -0.74_{-0.13}^{+0.70} \times L_{15}\\
&L_{LW3-0.6}^{DH} = 1.72_{-0.06}^{+0.02} \times L_{12}       -0.71_{-0.02}^{+0.04} \times L_{15}
\label{EQ:relat}
\end{split}
\end{equation}

Here again, as in Sect.~\ref{SEC:L127}, distant galaxies (light grey circles) follow the trend traced by local galaxies (filled triangles), hence confirming that no strong evidence is found for an evolution of the MIR SED of galaxies with redshift (see Fig.~\ref{FIG:10-15-1}). 
Again, distant galaxies exhibit a systematically lower $L_{10}/L_{15}$ ratio with two thirds of the distant LIRGs showing a lower ratio than local LIRGs. This cannot be explained by a selection effect since in this luminosity range, all galaxies exhibit ISOCAM flux densities above the 90\,\% completeness limit. Possible explanations are discussed in Sect.~\ref{SEC:discussion}.

\section{Discussion and conclusions}
\label{SEC:discussion}

We have presented a comparison of the mid infrared (MIR) and radio properties of local and distant galaxies (up to $z\sim$ 1.3) with particular emphasis on IR luminous galaxies, i.e. LIRGs and ULIRGs. Our goal was to assess the possibility of deriving a total IR luminosity, L(IR), corresponding to a star formation rate (SFR), on the sole basis of a measured flux density in the MIR for those galaxies. We derived three L(IR) estimates
using independantly the measured ISOCAM-15\,$\mu$m, MIPS-24\,$\mu$m and VLA-1.4 GHz flux densities and found that all three luminosities were consistent at the 40\,\% confidence level, i.e. an uncertainty similar to the one measured for the local MIR-FIR and radio-FIR correlations for IR luminous galaxies. This first result, obtained for a sample of 49 galaxies with redshifts up to $z\sim$ 1.3, and a sub-sample of 18 galaxies detected in the radio, suggests that the MIR can be used as a tracer of the SFR of distant actively star forming galaxies. However, the L(IR) computed from 15\,$\mu$m shows a 10-20\,\% larger
scatter than the one computed from 24\,$\mu$m
in comparison to the radio derived L(IR). This result suggests that the warm dust continuum and neutral PAH is a better tracer of the SFR than is the ionized PAH emission at rest-frame 7.7\,$\mu$m.

The redshift evolution of the 24 over 15\,$\mu$m colors show a bump around $z\sim$ 0.7 as one would expect from the entrance of PAH emission centered around 7.7\,$\mu$m into the 15\,$\mu$m passband. This behavior reinforces previous studies by \cite{Elbaz05} and \cite{Teplitz05} which were also interpreted as evidence for the presence of PAHs in distant IR luminous galaxies.

 We attempted to probe the same rest-frame wavelength range in local galaxies and high redshift galaxies.
We selected galaxies within two redshift bins centered at $z\sim$ 0.6 and 1, where the observed 15\,$\mu$m (24\,$\mu$m) passband corresponds to the rest-frame 10 and 7\,$\mu$m (15 and 12\,$\mu$m) respectively.  In both redshift bins, the distant galaxies fall in the same region as local galaxies of equivalent luminosities and provides an extension of local correlations to higher luminosities. We find marginal evidence of evolution with redshift of the MIR SEDs: two thirds of the distant galaxies exhibit rest-frame MIR colors (L(12\,$\mu$m)/L(7\,$\mu$m) and L(10\,$\mu$m)/L(15\,$\mu$m) luminosity ratios) below the median value measured for local galaxies. Possible explanations are examined below. It is indeed to be expected that an evolution should take place if not at the $z\sim$ 1 level, then at higher redshifts. The first cause that one may consider is the role played by metallicity. It was recently showed by \cite{engelbracht} and \cite{madden05} that the PAH over VSG continuum ratio was increasing as a function of metallicity, with a trend for galaxies with metallicities below $log(Z)$= 12$+log$(O/H)=8.2 to exhibit nearly no PAH contribution. \cite{Liang04} found the gas metallicity of LIRGs ranging between half-solar to solar ($log(Z)$=8.63--8.93) at similar redshifts. Hence, the influence of metallicity is not expected to be strong in our sample. Furthermore, it would be responsible for a stronger $L_{12}/L_{7}$ for the distant galaxies instead of the lower values that we tend to find. 
We measure a $L_{12}/L_{7}$ ratio for the distant LIRGs which is similar to the one found for "normal" nearby galaxies. This may suggest that distant LIRGs, which are probably more gas rich than local ones, could also be triggered by less violent causes, i.e. minor mergers, gas infall or even spiral waves instead of the major mergers found in local galaxies, as suggested by their morphologies \cite[see][]{bell05}. We note that the dust temperature measured in the far infrared also depends on the geometry of the regions of star formation \cite[][]{chanial05,Wang92}.
The lower $L_{10}/L_{15}$ ratio may result from a deeper 9.7\,$\mu$m silicate absorption.
Another possibility would be that the level of ionization of PAHs in distant galaxies is higher:  more ionised PAHs tend to present a lower $L_{12}/L_{7}$ \citep[e.g.][]{Dartois97}. Other causes might also play a role such as a different grain size distribution but it is not possible to disentangle between all those possibilities with existing data. Extended samples of objects at higher redshifts with Spitzer 16\,$\mu$m and 
24\,$\mu$m imaging as well as mid-infrared spectroscopy will soon be available which might help discriminate between the various effects.

We tested three commonly used libraries of template SEDs-- CE \citep{CE}, DH \citep{DH02} and LDP \citep{L04} -- in which the relative strengths of broad emission lines from PAHs with respect to warm dust continuum due to VSGs increases from CE to DH and to LDP. 

The LDP library predicts values for L(IR) which are 40\,\%  too low from the measured 15\,$\mu$m flux density and 30\,\% too large from the 24\,$\mu$m one, with respect to the radio ones. They also exhibit the largest dispersion and inconsistency between L(IR,15) and L(IR,24). This is mostly due to enhanced 8\,$\mu$m and decreased the continuum between 11-32 \,$\mu$m and  and to a lesser extent, due to a lack of evolution of MIR SEDs as a function of L(IR).

The CE and DH libraries both provide robust fits to both local and distant galaxies but we find evidence for more PAH emission in IR luminous galaxies than predicted in the CE SEDs. As a result, the CE library would overestimate L(IR) for distant galaxies by 20-30\,\%. If, as suggested above, distant galaxies do exhibit stronger PAH emission than local ones, then both libraries need to be revised and the CE library would overestimate L(IR) by up to 50\,\% for ULIRGs at $z\sim$ 1. Since LIRGs dominate the star formation rate density per unit comoving volume at least up to $z\sim$ 1.5, this evolution of MIR SEDs does not significantly affect the derivation of the cosmic star formation history (CSFH). We have checked this by replacing the CE SEDs by the DH ones in the \cite{CE} model and found two CSFH which are consistent within 30\,\%.

The true accuracy with which FIR luminosities can be determined at high redshift are being tested for the most luminous objects with Spitzer,
and will be pushed to fainter luminosities with the Herschel satellite scheduled for launch 2007, and the Atacama Large Millimeter Array, ALMA. Direct MIR spectroscopy with the InfraRed Spectrograph onboard Spitzer will also provide very useful constraints if pushed to detect distant LIRGs.

\acknowledgements{We wish to thank the anonymous referee for his constructive remarks which helped improving the paper in particular on the discussion of possible biases and Guilaine Lagache for useful comments. We also wish to thank Herv\'e Aussel for giving us the possibility to use his latest version of his ISOCAM-15\,$\mu$m catalog, including source deblending not included in the published version of the catalog.}

\bibliography{bib}

\begin{table*} [t]
\begin{turn}{90}
\begin{tabular}{ccccrrrrrrrrrrr} 
\hline
ID      & RA        & DEC           & z       &     f$_{15\,\mu{\rm m}}$                               & f$_{24\,\mu{\rm m}}$                    & f$_{8.5}$       &f$_{1.4}$  &  log L$_{\rm IR}^{\rm radio}$ & \multicolumn{3}{c}{log L(IR,24) [logL$_{\odot}$]} &  \multicolumn{3}{c}{log L(IR,15) [log L$_{\odot}$]}\\
        & (J2000)   &  (J2000)      &         &     $\mu$Jy                                            & $\mu$Jy                            & $\mu$Jy              & $\mu$Jy   &log [L${_\odot}$]              &      CE   &  DH   & LDP         & CE     & DH    & LDP    \\
(1)         &    (2)    &  (3)          &  (4)    & (5)                                      & (6)                       &   (7)            & (8)        &     (10)  & (11)  &(12)    &(13)    &(14)   &(15)&(16)   \\ 
\hline
 1& 189.1308136& 62.1871071& 1.013& $355^{+40}_{-40}$&  480$\pm$6    &--	&26.07	&11.63	&11.91	&11.98	&12.15	&12.35	&12.12	&11.87		\\
 2& 189.1436157& 62.2037354& 0.457& $448^{+60}_{-60}$& 1290$\pm$9    &39.20	&243.04	&11.78	&11.51	&11.56	&11.69	&11.33	&11.37	&11.41		\\
 3& 189.1438751& 62.2114830& 1.219& $363^{+60}_{-60}$&  446$\pm$5    &38.50	&178.62	&12.66	&12.24	&12.28	&12.43	&12.92	&12.50	&12.22		\\
 4& 189.1456909& 62.2068138& 0.562& $267^{+60}_{-60}$&  336$\pm$6    &--	&--	&--	&11.21	&11.22	&11.34	&11.27	&11.30	&11.22		\\
 5& 189.1536102& 62.1931000& 0.079& $300^{+60}_{-60}$&  732$\pm$9    &11.60	&--	&9.41	&9.53 &9.56	&9.48	&9.23	&9.20	&9.22		\\
 6& 189.1663666& 62.2139931& 0.846& $295^{+60}_{-60}$&  493$\pm$6    &--	&55.41	&11.77	&11.69	&11.76	&11.94	&11.86	&11.74	&11.57		\\
 7& 189.1924133& 62.1951141& 1.016& $125^{+41}_{-41}$&  290$\pm$5    &--	&21.04	&11.54	&11.67	&11.73	&11.88	&11.60	&11.51	&11.26		\\
 8& 189.1926117& 62.2576752& 0.851& $418^{+52}_{-52}$&  544$\pm$8    &--	&40.94	&11.65	&11.75	&11.83	&11.98	&12.09	&11.95	&11.78		\\
 9& 189.2012482& 62.2408028& 0.139& $307^{+48}_{-48}$&  460$\pm$6    &--	&--	&--	&9.88	&9.88	&9.78	&9.71	&9.76	&9.72		\\
10& 189.2062225& 62.2353172& 0.752& $150^{+35}_{-35}$&  186$\pm$6    &--	&--	&--	&11.18	&11.22	&11.33	&11.27	&11.23	&11.04		\\
11& 189.2102509& 62.2212448& 0.851& $157^{+45}_{-45}$&  196$\pm$6    &--	&--	&--	&11.27	&11.34	&11.45	&11.48	&11.42	&11.22		\\
12& 189.2071533& 62.2203903& 0.474& $115^{+40}_{-40}$&  371$\pm$10   &14.00	&48.95	&11.13	&11.07	&11.06	&11.14	&10.56	&10.73	&10.64		\\
13& 189.2129822& 62.1753387& 0.410& $341^{+42}_{-42}$&  984$\pm$9    &14.00	&94.20	&11.26	&11.28	&11.31	&11.41	&11.04	&11.10	&11.12		\\
14& 189.2157745& 62.2317200& 0.556& $151^{+36}_{-36}$&  203$\pm$6    &--	&--	&--	&11.00	&10.99	&11.06	&10.93	&11.00	&10.88		\\
15& 189.2217255& 62.1880875& 0.935& $174^{+33}_{-33}$&  367$\pm$6    &--	&--	&--	&11.65	&11.72	&11.87	&11.68	&11.61	&11.35		\\
16& 189.2223816& 62.1944351& 1.270& $180^{+36}_{-36}$&  322$\pm$6    &12.60	&73.97	&12.32	&12.17	&12.22	&12.35	&12.57	&12.23	&11.95		\\
17& 189.2245483& 62.2151070& 0.642& $179^{+31}_{-31}$&  200$\pm$6    &--	&--	&--	&11.11	&11.12	&11.22	&11.19	&11.17	&11.03		\\
18& 189.2407684& 62.2486877& 0.849& $225^{+23}_{-23}$&  366$\pm$8    &9.40	&--	&11.63	&11.57	&11.63	&11.75	&11.69	&11.63	&11.43		\\
19& 189.2457886& 62.2025566& 0.852& $157^{+19}_{-19}$&  269$\pm$6    &--	&--	&--	&11.42	&11.47	&11.58	&11.48	&11.42	&11.22		\\
20& 189.2496338& 62.2472992& 0.761& $295^{+43}_{-43}$&  466$\pm$6    &--	&41.97	&11.54	&11.59	&11.66	&11.83	&11.69	&11.64	&11.46		\\
21& 189.2744446& 62.1983376& 0.903& $431^{+34}_{-34}$&  655$\pm$8    &--	&57.63	&11.86	&11.89	&11.96	&12.12	&12.26	&12.02	&11.86		\\
22& 189.1266937& 62.2024765& 0.457& $145^{+48}_{-48}$&   95$\pm$5    &--	&--	&--	&10.51	&10.44	&10.42	&10.65	&10.80	&10.72		\\
23& 189.1399536& 62.2222900& 0.845& $122^{+40}_{-40}$&  323$\pm$7    &--	&--	&--	&11.50	&11.58	&11.69	&11.31	&11.30	&11.04		\\
24& 189.1552277& 62.2263107& 0.847& $184^{+50}_{-50}$&  106$\pm$5    &--	&--	&--	&10.97	&11.04	&11.11	&11.57	&11.48	&11.31		\\
25& 189.1619263& 62.2160149& 1.143& $155^{+50}_{-50}$&  244$\pm$7    &--	&--	&--	&11.79	&11.87	&11.99	&12.08	&11.91	&11.64		\\
26& 189.1758118& 62.2627182& 0.857& $459^{+60}_{-60}$&  850$\pm$8    &22.70	&126.45	&12.14	&11.96	&12.03	&12.21	&12.21	&12.00	&11.85		\\
27& 189.1788483& 62.2046394& 0.454& $49^{+36}_{-09} $&  114$\pm$5    &--	&--	&--	&10.56	&10.51	&10.48	&10.10	&10.24	&10.10		\\
28& 189.1858521& 62.2180252& 0.484& $33^{+11}_{-15} $&  108$\pm$5    &--	&--	&--	&10.56	&10.53	&10.50	&9.98	&10.12	&9.97		\\
29& 189.2199860& 62.2483215& 0.089& $42^{+29}_{-9}  $&   32$\pm$7    &--	&--	&--	&8.37	&8.38	&8.29	&8.47	&8.36	&8.50		\\
30& 189.2266083& 62.2429657& 0.577& $51^{+75}_{-51} $&  104$\pm$6    &--	&--	&--	&10.73	&10.73	&10.77	&10.30	&10.46	&10.24		\\
31& 189.2276459& 62.1909523& 0.254& $42^{+29}_{-09} $&  173$\pm$6    &--	&--	&--	&10.10	&10.05	&9.95	&9.45	&9.47	&9.46		\\
32& 189.2387543& 62.2166634& 0.474& $60^{+40}_{-23} $&   30$\pm$7    &--	&--	&--	&10.16	&9.99	&9.93	&10.21	&10.40	&10.24		\\
33& 189.2483521& 62.1982346& 1.020& $48^{+56}_{-10} $&  117$\pm$6    &--	&--	&--	&11.24	&11.31	&11.40	&10.98	&11.05	&10.70		\\
34& 189.2680359& 62.2462730& 2.211& $80^{+74}_{-19} $&  376$\pm$4    &--	&--	&--	&13.32	&12.90	&12.66	&13.47	&13.32	&12.81		\\
35& 189.2693481& 62.2415276& 0.560& $72^{+65}_{-16} $&  173$\pm$5    &--	&26.45	&11.03	&10.96	&10.92	&11.00	&10.44	&10.59	&10.44		\\
36& 189.2847595& 62.2147064& 0.838& $115^{+11}_{-11}$&   81$\pm$6    &--	&--	&--	&10.83	&10.94	&10.98	&11.25	&11.22	&11.01		\\
37& 189.2052155& 62.1968956& 0.960& $75^{+71}_{-19} $&   86$\pm$6    &--	&--	&--	&10.97	&11.06	&11.12	&11.18	&11.16	&10.89		\\
\hline																	      
\end{tabular} 					      					     						      
\end{turn}					      					     
\caption{Description of the sample. The ISOCAM (f$_{15\,\mu{\rm m}}$) and MIPS (f$_{24\,\mu{\rm m}}$) flux densities are given with their 68\,\% error bars. f$_{1.4}$ and f$_{8.5}$ are the 1.4 and 8.5 GHz flux densities respectively from Morrison et al. (in prep.) and Richards(2000). The logarithm of the total IR luminosities were derived using the recipe described in the paper. (continuing in Table~\ref{TAB:sample2}).} 
%
\label{TAB:sample1}					      					     						      
\end{table*}																            	\clearpage
															      
\begin{table}																      
\begin{turn}{90}															      
\begin{tabular}{ccccrrrrrrrrrrr} 													      
\hline
ID      & RA        & DEC            & z        &     f$_{15\,\mu{\rm m}}$                               & f$_{24\,\mu{\rm m}}$                    & f$_{8.5}$       &f$_{1.4}$  &  log L$_{\rm IR}^{\rm radio}$ & \multicolumn{3}{c}{log L(IR,24) [log L$_{\odot}$]} &  \multicolumn{3}{c}{log L(IR,15) [log L$_{\odot}$]}\\
        & (J2000)   &  (J2000)       &          &     $\mu$Jy                                            & $\mu$Jy                                 & $\mu$Jy         & $\mu$Jy   &[log L${_\odot}$]&      CE   &  DH   & LDP         & CE     & DH    & LDP    \\
(1)     &    (2)    &  (3)           &  (4)     & (5)                                                    & (6)                                     &   (7)           & (8)       &     (10)  & (11)  &(12)    &(13)    &(14)   &(15)&(16)   \\ 
\hline
38& 189.1941833& 62.1803780& 0.940& $295^{+40}_{-40}$&  354$\pm$7    &--	&47.40	&11.81	&11.64	&11.71	&11.87	&12.04	&11.90	&11.67		\\
39& 189.1533661& 62.2037392& 0.848& $202^{+50}_{-50}$&  379$\pm$5    &--	&--	&--	&11.58	&11.65	&11.79	&11.62	&11.58	&11.35		\\
40& 189.2614288& 62.2338943& 1.246& $144^{+20}_{-20}$&  334$\pm$8    &--	&30.39	&11.91	&12.11	&12.19	&12.28	&12.33	&12.04	&11.77		\\
41& 189.1583710& 62.1882019& 1.017& $212^{+55}_{-55}$&  291$\pm$5    &--	&--	&--	&11.67	&11.74	&11.89	&11.97	&11.85	&11.58		\\
42& 189.2758789& 62.2258720& 0.752& $86^{+78}_{-20} $&  138$\pm$7    &--	&--	&--	&11.05	&11.08	&11.16	&10.91	&10.94	&10.71		\\
43& 189.1595612& 62.1975403& 0.841& $212^{+55}_{-55}$&  230$\pm$3    &--	&--	&--	&11.34	&11.40	&11.54	&11.64	&11.59	&11.37		\\
44& 189.1715393& 62.2392883& 0.518& $76^{+71}_{-19} $&   50$\pm$6    &--	&--	&--	&10.42	&10.29	&10.25	&10.38	&10.56	&10.44		\\
45& 189.1748047& 62.2015877& 0.432& $52^{+34}_{-09} $&   83$\pm$5    &--	&--	&--	&10.40	&10.31	&10.25	&10.08	&10.22	&10.09		\\
46& 189.2253876& 62.2318687& 0.850& $47^{+31}_{-09} $&   39$\pm$5    &--	&--	&--	&10.42	&10.56	&10.55	&10.63	&10.79	&10.48		\\
47& 189.2584381& 62.2231522& 0.409& $83^{+76}_{-19} $&   46$\pm$7    &--	&--	&--	&10.14	&10.00	&9.927	&10.22	&10.40	&10.27		\\
48& 189.1679382& 62.2253952& 0.483& $36^{+11}_{-15} $&  135$\pm$6    &--	&--	&--	&10.59	&10.60	&10.58	&10.01	&10.16	&10.00		\\
49& 189.1799927& 62.1967506& 1.007& $48^{+57}_{-10} $&  113$\pm$5    &--	&--	&--	&11.20	&11.28	&11.35	&10.96	&11.03	&10.69		\\
\hline																	       
\end{tabular} 					      				
\end{turn}					      					     
\caption {Continuation of Table~\ref{TAB:sample1}.}  					      					     
\label{TAB:sample2}					      					     
\end{table} 

\end{document}